# Factors Enabling Delocalized Charge-Carriers in Pnictogen-Based Solar Absorbers: In-depth Investigation into CuSbSe$_2$

Yuchen Fu,[1,†] Hugh Lohan,[1,2,†] Marcello Righetto,[3] Yi-Teng Huang,[1,4] Seán R. Kavanagh,[2] Chang-Woo Cho,[5] Szymon J. Zelewski,[4,6] Young Won Woo,[2,7] Harry Demetriou,[2] Martyn A. McLachlan,[2] Sandrine Heutz,[2,8] Benjamin A. Piot,[5] David O. Scanlon,[9] Akshay Rao,[4] Laura M. Herz,[3,10] Aron Walsh,[2] and Robert L. Z. Hoye[1,*]

1.  Inorganic Chemistry Laboratory, University of Oxford, South Parks Road, Oxford OX1 3QR, United Kingdom

2.  Department of Materials and Centre for Processable Electronics, Imperial College London, Exhibition Road, London SW7 2AZ, United Kingdom

3.  Department of Physics, University of Oxford, Clarendon Laboratory, Parks Road, Oxford, OX1 3PU, United Kingdom

4.  Cavendish Laboratory, University of Cambridge, JJ Thomson Ave, Cambridge CB3 0HE, United Kingdom

5.  Laboratoire National des Champs Magnétiques Intenses, CNRS, LNCMI, Université Grenoble Alpes, Université Toulouse 3, INSA Toulouse, EMFL, F-38042 Grenoble, France

6. Department of Experimental Physics, Faculty of Fundamental Problems of Technology, Wrocław University of Science and Technology, Wybrzeże Wyspiańskiego 27, 50-370 Wrocław, Poland

7. Department of Materials Science and Engineering, Yonsei University, Seoul 03722, Korea

8. London Centre for Nanotechnology, Imperial College London, Prince Consort Road, London, SW7 2AZ, United Kingdom

9. School of Chemistry, University of Birmingham, Birmingham, B15 2TT, United Kingdom

10. Institute for Advanced Study, Technical University of Munich, Lichtenbergstrasse 2a, D-85748 Garching, Germany

[†]These authors contributed equally to this work.

**Email:** robert.hoye@chem.ox.ac.uk (R.L.Z.H.)

**Abstract**

Inorganic semiconductors based on heavy pnictogen cations ($Sb^{3+}$ and $Bi^{3+}$) have gained significant attention as potential nontoxic and stable alternatives to lead-halide perovskites for solar cell applications. A limitation of these novel materials, which is being increasingly commonly found, is carrier localization, which substantially reduces mobilities and diffusion lengths. Herein, the layered příbramite $CuSbSe_2$ is investigated and discovered to have delocalized free carriers, as shown through optical pump terahertz probe spectroscopy and temperature-dependent mobility measurements. Using a combination of theory and experiment, it is found that the underlying factors are: 1) weak coupling to acoustic phonons due to low deformation potentials, as lattice distortions are primarily accommodated through rigid inter-layer movement rather than straining inter-atomic bonds, and 2) weak coupling to optical phonons due to the ionic contributions to the dielectric constant being low compared to electronic contributions. This work provides important insights into how pnictogen-based semiconductors avoiding carrier localization could be identified.

**Keywords**: Carrier-Phonon Coupling, Carrier Localization, Perovskite-Inspired Materials, Chalcogenides, Earth-Abundant Solar Absorbers, Ultrafast Spectroscopy, Solution Processing

**Introduction**

Semiconductors based on heavy pnictogens (namely $Sb^{3+}$ and $Bi^{3+}$) have gained a surge of interest over the past few years because of their potential to replicate the defect tolerance of lead-halide perovskites (LHPs), whilst overcoming their toxicity and stability limitations[1-6]. Defect tolerance is the ability of semiconductors to achieve low non-radiative recombination rates and maintain high mobilities despite high defect densities, and it is believed that such an effect occurs in LHPs in part because of its unusual electronic structure at its band-edges, which comes about from the strong contributions of the Pb $6s^2$ electrons to the valence band density of states[7]. As such, there has been a focus on compounds based on heavy post-transition metal cations $In^+$, $Sn^{2+}$, $Sb^{3+}$ and $Bi^{3+}$, which have valence $ns^2$ electrons and, unlike Pb, are fully compliant with regulations on elements that can be safely used in consumer electronics[8]. Heavier cations are preferable, as spin-orbit coupling increases with effective nuclear charge, which results in a smaller bandgap, thereby increasing the chances of dominant defects forming shallow traps[7]. Among these four cations, $Sb^{3+}$ and $Bi^{3+}$ are especially promising because they are not severely limited in supply or expensive (unlike In)[9], and their valence $ns^2$ electron pair is stable (unlike $Sn^{2+}$)[2]. Indeed, many Sb- and Bi-based inorganic semiconductors have been found to be more environmentally and thermally stable than LHPs[10-16], and have also been found to avoid the self-doping that is prevalent in Sn perovskites[17-20].

Early work developing solar absorbers from these heavy pnictogen-based compounds focussed primarily on their charge-carrier lifetimes, in addition to their bandgaps and absorption coefficients, with the assumption that there was no significant difference in the mobilities between these materials[1]. Surprisingly, some Bi-based thin film materials were found to exhibit lifetimes in the hundreds of nanoseconds to microseconds range[12,21,22], far exceeding the minority carrier lifetimes of conventional inorganic semiconductors (1-10 ns) or LHPs (~100 ns in polycrystalline thin films)[23]. Recently, it was realized that this slow long-time decay in the population of photogenerated charge-carriers came about from carrier localization, in which the wavefunctions of charge-carriers or excitons are confined to within a unit cell or smaller, leading to the formation of small polarons or self-trapped excitons[24,25]. Carrier localization substantially reduces mobilities and therefore limits diffusion lengths, despite the slow decay in the population of excitations[12,24,25]. For example, although $Cs_2AgBiBr_6$ halide elpasolites have photogenerated charge-carriers decaying with a time-constant in the microsecond range, steady-state mobilities only reach up to ~10 $cm^2$ $V^{-1}$ $s^{-1}$ in single crystals[26]. Electron diffusion lengths as short as 30 nm have been found in polycrystalline $Cs_2AgBiBr_6$, which partly accounts for the low photovoltaic power conversion efficiencies (PCEs) that are usually well below 4%[27]. Recent investigations into the wider family of bismuth-halide and bismuth-chalcogenide semiconductors have found carrier localization to be so prevalent that it is being described as a hallmark of these materials[10-12,14,28-31]. The effect of carrier localization on Sb-based compounds is not as well established. One of the best-studied of these materials is the antimony chalcogenide family of compounds ($Sb_2S_3$ and $Sb_2Se_3$). There are currently strong disagreements in the community regarding whether self-trapping occurs in these materials, limiting open-circuit voltages up to a maximum of 0.8 V[30,32-34], or whether the performance is instead limited by charged defects[35-37]. In $Cs_2AgSbBr_6$, on the other hand, charge-carrier localization proceeds on a picosecond timescale, similar to that in $Cs_2AgBiBr_6$, with alloying of the two materials exacerbating such effects, owing to localized charge-carriers being more susceptible to energetic disorder[38].

It is clear that the future development of pnictogen-based perovskite-inspired materials for optoelectronic devices urgently requires not only consideration of defects, but also insights into how charge-carrier localization may be avoided in these materials. Very recently, we provided hints in this direction with detailed spectroscopic and computational investigations into bismuth oxyiodide (BiOI)[39,40]. BiOI exhibits a red-shift in the photoluminescence (PL) spectrum compared to its optical bandgap, which would typically be considered to arise from self-trapping. However, we found that this red-shift can be fully accounted for by the coupling between charge-carriers and two longitudinal optical (LO) phonon modes generated coherently through photo-excitation. The delocalized nature of these large polarons was verified from computations of the wavefunction of the lowest-energy exciton, as well as magneto-optical spectroscopy measurements. The mobilities reached up to 83 cm$^2$ V$^{-1}$ s$^{-1}$ in the in-plane direction, exceeding the mobilities of self-trapped carriers (typically ~10 cm$^2$ V$^{-1}$ s$^{-1}$ or lower)[10-12,24-26,28]. In a separate recent work, we showed through optical pump terahertz probe (OPTP) spectroscopy measurements of thin film samples that BiOI avoids charge-carrier localization in both the in-plane and out-of-plane directions[40]. Therefore unlike most novel bismuth-halide semiconductors, BiOI exhibits band-like transport, and we speculated that this was a result of its layered nature and the large thickness of each layer, resulting in an electronic dimensionality higher than 2D[39]. However, the detailed mechanisms, as well as the role of acoustic phonons and how they interact with charge-carriers, are yet to be determined.

Inspired by this recent work, herein we investigate a related layered Sb-based compound, CuSbSe$_2$. This material is a příbramite, which is the Se analogue to the chalcostibite CuSbS$_2$. More broadly, the ABZ$_2$ family of materials (A = monovalent cation, B = $Sb^{3+}$ or $Bi^{3+}$, Z = chalcogen) have gained attention as promising pnictogen-based semiconductors. This is because $AgBiS_2$ photovoltaics recently reached a certified PCE of 8.85%[41], which is among the highest for any Bi-based solar absorber. Both $AgBiS_2$, and the related $NaBiS_2$ compound, were found to be stable in air and have slow decays in their photoexcited charge-carriers[12,13,41]. Our detailed investigations into $NaBiS_2$ showed that this was due to carrier localization, which was facilitated by localized S 3p states that form in regions where there are high coordination of Na around S, likely capturing holes and leading to the formation of small hole polarons[12]. Recently, the presence of charge-carrier localization in $AgBiS_2$ was also reported, and the degree of charge-carrier localization can be tuned through cation disorder engineering[42]. $CuSbSe_2$ avoids the cation disorder found in both $NaBiS_2$ and $AgBiS_2$ owing to the $Cu^+$ and $Sb^{3+}$ cations having sufficiently different radii (60 pm and 76 pm, respectively)[43], as well as the stereochemical activity of the lone pair on $Sb^{3+}$, such that the smaller $Cu^+$ occupies tetrahedral sites, while $Sb^{3+}$ occupies trigonal pyramidal sites (which allows the lone pair on $Sb^{3+}$ to be projected out into space). Furthermore, the thickness of each layer in $CuSbSe_2$ (5.70 Å) is comparable to that of BiOI (6.14 Å)[39]. There is therefore a possibility that $CuSbSe_2$ may be able to avoid the charge-carrier localization found in $NaBiS_2$, $AgBiS_2$ and most Bi-halide compounds, and if so, the mechanism by which this occurs will be of paramount importance to learn how delocalized excitations can be achieved more broadly across the wider family of pnictogen-based perovskite-inspired materials.

In this work, we developed a solution processing route to achieve phase-pure $CuSbSe_2$ thin films. The optical phonon modes present were determined through Raman and infrared (IR) spectroscopy, and the nature of excitations (*i.e.*, whether free charge-carriers or excitons formed) was determined through Elliott model fitting of the optical absorption spectra, and correlated with computations of the exciton binding energy. To understand whether these excitations are localized, the charge-carrier kinetics were measured by transient absorption spectroscopy (using a femtosecond pulsed excitation laser), and photoconductivity transients by OPTP spectroscopy, along with measurements of temperature-dependent mobility. The underlying mechanisms behind the nature of excitations were established through calculations of the strength of coupling with acoustic phonons (acoustic coupling constant) and LO phonons (Fröhlich coupling constant), as well as calculations of the key parameters that influence these coupling constants, namely the deformation potential, dielectric tensor, bonding/antibonding nature of crystal orbitals at the band extrema, changes in bond lengths and interlayer spacing arising from distortions, as well as the Born effective charge. The understanding gained from investigating the case of $CuSbSe_2$ can provide insights into how we could design heavy pnictogen-based semiconductors with band-like transport, which will be critical for creating more promising earth-abundant solar absorbers.

**Results**

***Structure, synthesis and vibrational properties of CuSbSe₂ thin films***

CuSbSe$_2$ has a layered structure (Fig. 1a), with an orthorhombic unit cell that is similar to that of chalcostibites (*Pnma* space group)[44,45]. The CuSbSe$_2$ layers are held together by *van der Waals* interactions. Each Sb atom is bonded to three Se atoms in a trigonal pyramidal geometry, while each Cu atom is bonded with four Se atoms in a tetrahedral arrangement. The CuSe$_4$ tetrahedra and SbSe$_3$ trigonal pyramids share corners (of Se). The distance between Sb and Se atoms separated by the interlayer gap (3.26 Å) is too large to form covalent bonds between these atoms. From the structure shown in Fig. 1a, it can be seen that there is static stereochemical activity of the 5s$^2$ lone pair on Sb$^{3+}$ [46-61]. This stereochemical activity is also found in CuSbS$_2$[62-64], and occurs because the Sb 5s and chalcogen valence p orbitals are close enough in energy for mixing, such that a second order Jahn-Teller distortion can occur[65]. Indeed, the Sb 5s lone pair is also stereochemically active in Sb$_2$Se$_3$[66,67], indicating that the orbital energy levels of the Sb 5$s$ and Se 4$p$ states are close enough in energy to interact.

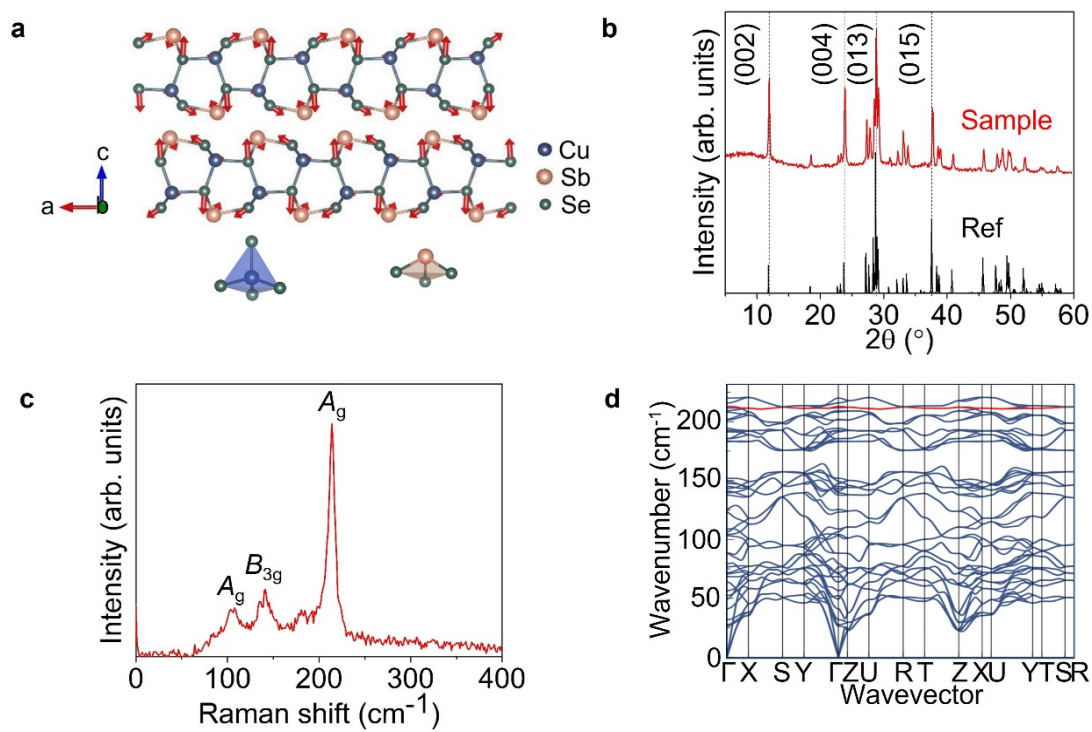

**Fig. 1 | Structural and phonon properties of CuSbSe₂. a,** Crystal structure of CuSbSe₂, viewed along the *b* axis, and with the dominant $A_g$ Raman mode shown in red arrows. The bonding environments of Cu and Sb are illustrated below the crystal structure. **b,** X-ray diffraction (XRD) pattern of solution-processed thin films compared with the reference pattern of CuSbSe₂ (ICSD database, ID 418754; detailed fitting in Supplementary Fig. 1a). **c,** Raman spectrum of spin-coated CuSbSe₂ thin film with phonon modes of the most intense peaks labelled. **d,** Phonon dispersion curve of CuSbSe₂. The band containing the dominant $A_g$ mode is highlighted in red.

Previous efforts at growing CuSbSe₂ focussed on vacuum-based approaches (*e.g.*, sputter deposition[68], close-space sublimation[69]), methods that have long reaction times (*e.g.*, fusion method[70] or selenization of metal precursors[71]), or processes involving the use of toxic precursors (*e.g.*, hydrazine solvent[72,73]). Solution-processing is advantageous in requiring less capital-intensive equipment than vacuum-based processing[74-76], but at the same time, it is critical to avoid the use of toxic solvents[77].

More recently, a more benign solvent system than hydrazine, comprised of a thiol-amine mixture, has been found to be effective in dissolving chalcogenide precursors[78-80] and successfully used to deposit absorber layers of photovoltaic devices, such as $Cu_2ZnSn(S,Se)_4$[81,82], $Cu(In, Ga)Se_2$[83,84] and $CuIn(S, Se)_2$[85]. In this work, we develop a thiol-amine route to synthesize $CuSbSe_2$ thin films for the first time, as detailed in Methods. To achieve crystalline films, we dried the films at 100 °C for 2 min in a $N_2$-filled glovebox, before crystallizing at 400 °C for 2 min in a tube furnace filled with Ar (~1200 mTorr pressure). Pawley fitting of the X-ray diffraction (XRD) pattern of these films with the reference pattern for $CuSbSe_2$ (ICSD database, coll. code 418754) showed that all measured peaks were accounted for by the príbramite phase (Fig. 1b, details in Supplementary Fig. 1).

To confirm the phase purity of the $CuSbSe_2$ films, Raman and Fourier transform infrared (FTIR) spectroscopy were employed. For the *Pnma* space group ($D_{2h}^{16}$), there are four Raman active mode symmetries ($A_g$, $B_{1g}$, $B_{2g}$, and $B_{3g}$), along with three IR active mode symmetries ($B_{1u}$, $B_{2u}$, and $B_{3u}$)[72,86]. In the Raman spectrum measured from the thin film samples (Fig. 1c), three obvious peaks at $105.7\pm0.2$ cm$^{-1}$ ($A_g$), $141.7\pm0.6$ cm$^{-1}$ ($B_{3g}$) and $213.7\pm0.2$ cm$^{-1}$ ($A_g$) can be observed, and all of them have been reported to come from the príbramite phase of $CuSbSe_2$[68,87-89]. The results agree with our calculated phonon spectrum for CuSbSe$_2$ (Fig. 1d and Supplementary Table 2). In the FTIR spectrum (Supplementary Fig. 2), two relatively strong peaks centred at 182.8$\pm$0.2 cm$^{-1}$ and 223.1$\pm$0.1 cm$^{-1}$ were observed. According to our calculations, these two peaks can be assigned to the $B_{2u}$ and $B_{3u}$ modes, respectively. Combining these XRD, Raman and FTIR measurements, we can conclude that the spin-coated CuSbSe$_2$ thin films are phase-pure after annealing at 400 °C for 2 min.

It is worth noting that the intensity of the $A_g$ mode at $\approx$213 cm$^{-1}$ is much higher than other Raman-active modes (Fig. 1c). We also constructed the phonon dispersion curve (Fig. 1d), and the band of the dominant $A_g$ mode at 213.7$\pm$0.2 cm$^{-1}$ is highlighted. The vibrations associated with this mode are calculated and illustrated by the red arrows in Fig. 1a, showing this to be an intra-layer breathing mode. The phonon density of states is shown in Supplementary Fig. 2 and compared with the FTIR and Raman spectra.

***Optoelectronic properties of CuSbSe$_2$***

Having developed phase-pure samples and understood the dominant phonon modes in CuSbSe$_2$, we next needed to understand the nature of excitations and their kinetics. The black solid line in Fig. 2a shows the measured optical absorbance curve of CuSbSe$_2$ with the fit (red dashed line) obtained from Elliott's theory[90], following a previously-reported procedure[91]. We note that despite a significant lineshape broadening ($\Gamma \sim 90$ meV), the fit matches with the measured spectrum well. The deconvolution of the excitonic and continuum contributions yields a weak and broad excitonic contribution, described by an exciton binding energy ($E_b$) of 9$\pm$4 meV. This matches well with the density functional theory (DFT) calculations we made on $CuSbSe_2$, from which we obtained an $E_b$ of 8.7 meV, as estimated using the Wannier-Mott hydrogenic model[92]. Given that these $E_b$ values are well below $kT$ at room temperature (26 meV), we would expect $CuSbSe_2$ to predominantly exhibit free charge-carriers rather than bound excitons.

We also note that the absorbance curve shows a shoulder at approximately 1.4 eV (details in Supplementary Fig. 3a), which could either arise from excitons or from the electronic structure of the material. To distinguish between these two possibilities, we computed the optical absorption spectrum (Supplementary Fig. 3b) from the frequency-dependent dielectric tensor using hybrid DFT (HSE06 functional)[93]. The computed absorption spectrum reproduced the experimentally-observed shoulder in absorption. Our calculations were carried out in the independent particle approximation[94], and therefore phonon-assisted transitions and polaronic/excitonic effects were not considered. This analysis shows that the shoulder in the absorption spectrum of CuSbSe₂ arises because of its electronic structure.

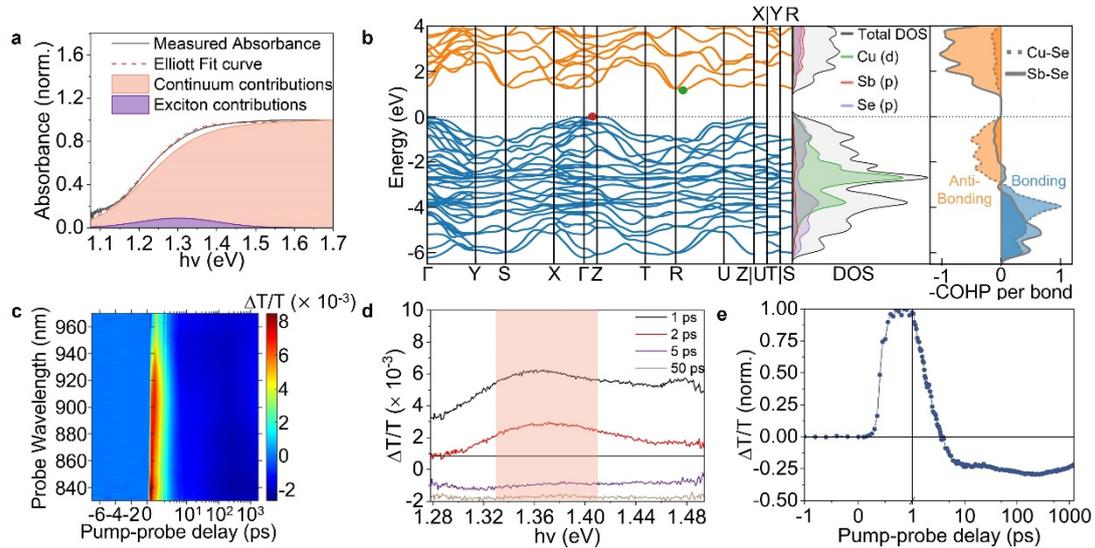

**Fig. 2 | Optical and electronic properties of CuSbSe₂. a,** Comparison between the measured optical absorbance curve (black solid line) and fit with the Elliott model (red dashed line). The contributions from the exciton and continuum to the optical absorption spectrum are represented by the areas shaded purple and pink, respectively. **b,** Electronic band structure of CuSbSe₂ (left panel; the highest occupied state set to 0 eV), along with electronic density of states curves (middle panel), and crystal orbital Hamilton population (COHP) diagram (right panel). The bonding and anti-bonding interactions are represented by blue and orange, respectively. **c,** Short-time transient absorption (TA) signal colour map of CuSbSe₂ films excited by a 800 nm wavelength pump (150 fs pulse width, 41 μJ cm⁻² pulse⁻¹ fluence, with 500 Hz repetition rate), along with **d,** short-time TA spectra for pump-probe delays of 1, 2, 5 and 50 ps, and **e,** its normalized ground state bleach (GSB) signal kinetics. The GSB kinetics were acquired by averaging the signals from 1.33 to 1.41 eV (pink shaded area in **d**) and normalized to the maximum $\Delta T/T$ value.

To understand the kinetics of the free charge-carriers in CuSbSe₂, short-time transient absorption (TA) spectroscopy was employed. In short-time TA measurements, the sample was excited with the 800 nm wavelength pulsed laser. After excitation, probe pulses, comprising a broadband near-IR spectrum, were used to measure the relative change in transmittance ($\Delta T/T$) of the sample at certain delays after pumping, with pump-probe delays ranging from 1 to 1000 ps. The positive ground state bleach (GSB) signal on a $\Delta T/T$ scale is usually proportional to the photo-excited carrier population near the band edges. The decay in the GSB signal therefore reflects the depopulation of charge-carriers near the band edges. Meanwhile, negative photo-induced absorption (PIA) can also occur, and the possible origins of PIA include self-trapping, absorption related to defect states or excitation to higher energy states. Strong PIA signals can interfere with the GSB signals. The results of short-time TA measurements are shown in Fig. 2c-e. In the short-time TA spectrum (Fig. 2d), we can observe a broadband GSB signal centred at approximately 1.36 eV. However, the positive GSB signal was pulled down by a strong PIA signal within 5-10 ps, which can also be observed in the normalized TA signal kinetics (Fig. 2e). The strong PIA signal makes it difficult to estimate the charge-carrier lifetime of $CuSbSe_2$ films. As for long-time TA measurements (355 nm wavelength pump, with pump-probe delays ranging from 1 to 1000 ns), the GSB signal was completely suppressed by a PIA signal and no GSB signal could be observed (Supplementary Fig. 4a-c). The strong PIA signal in TA measurements made it necessary to use other techniques to better understand charge-carrier kinetics in $CuSbSe_2$. Nevertheless, the breadth of the GSB observed in the short-time TA measurements, along with the absence of PIA on either side initially is consistent with these excitations originating from charge-carriers rather than excitons.

### *Experimental investigation into charge-carrier-phonon coupling in CuSbSe₂*

To gain more in-depth insights into the nature of the excitations in $CuSbSe_2$, we employed optical pump terahertz probe (OPTP) measurements. The fractional change in transmitted terahertz (THz) field amplitude $-\Delta T/T$ is monitored in OPTP measurements with sub-picosecond time resolution following a 400-nm wavelength pulsed excitation (see Methods for details). The measured $-\Delta T/T$ signal is proportional to the photoconductivity $\Delta\sigma$ of the studied thin film, making it ideal for investigating charge-carrier localization processes. As demonstrated for several Bi-based semiconductors, the charge-carrier localization process yields a photoconductivity decay on the sub-picosecond timescale[10,12,28], as a result of the lower mobility of localized charge-carriers. By comparison, defect-assisted trapping would cause a slower decay in the photoconductivity because charge-carriers need to diffuse to the defect states before they are trapped (reducing mobility), or undergo non-radiative recombination (reducing the photoexcited charge-carrier population)[10]. The different timescales of photoconductivity decay can provide insights into the trapping mechanisms inside materials, especially since free charge-carriers rather than excitons form in CuSbSe$_2$ (see end of Supplementary Note 3 for details).

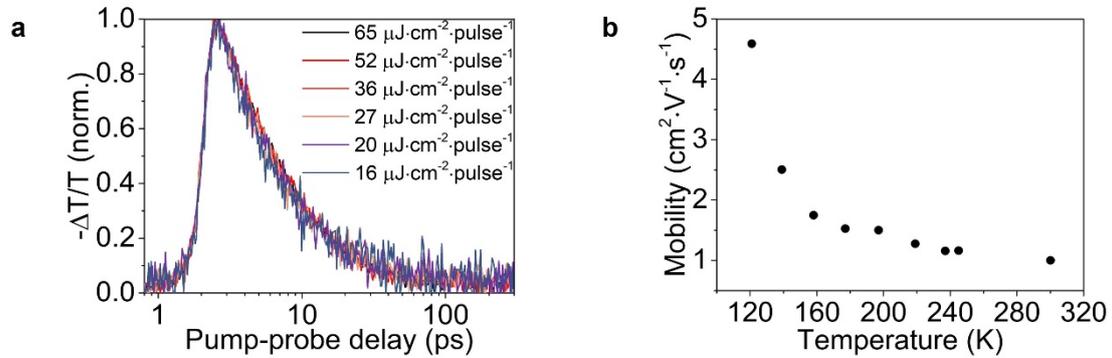

**Fig. 3 | Spectroscopic and temperature-dependent studies on carrier-phonon coupling in CuSbSe$_2$. a,** Normalized comparison between fluence-dependent optical pump terahertz probe (OPTP) transients measured for CuSbSe$_2$ thin films following 400 nm wavelength pulsed excitation. **b,** Temperature-dependent mobility of CuSbSe$_2$ thin films determined using Hall effect measurements.

In the case of CuSbSe$_2$, we found that 92% of the original OPTP signal was lost over a

period of 50 ps. This is a slower decay than observed from short-time TA measurements

(Fig. 2e), likely because the GSB kinetics had the rise in PIA superimposed upon it,

obscuring the real decay of the photogenerated charge-carriers. The decay observed in

the OPTP measurements is significantly slower than previously reported for other

bismuth-based semiconductors which have charge-carrier dynamics dominated by

localization processes. The localization timescales of Cs$_2$AgBiBr$_6$, Cu$_2$AgBiI$_6$ and

NaBiS$_2$ were reported to range from <1 ps to a few ps [10-12,28]. The OPTP kinetics for

CuSbSe$_2$ are therefore consistent with delocalized free charge-carriers diffusing to defect states and undergoing non-radiative recombination, rather than undergoing carrier localization.

To verify that delocalized large polarons instead of small polarons form in $CuSbSe_2$, we measured the temperature dependence of the mobility through Hall effect measurements. The mobility of large polarons decreases as the temperature increases because increasing temperature can lead to more phonons that charge-carriers can be coupled to, and we confirmed this in our computations of the temperature-dependent mobility of charge-carriers scattered by LO phonons (Supplementary Fig. 5) [25,95-97]. On the contrary, small polarons show an increase in the mobility with increasing temperature since small polarons can only hop between lattice sites, and increasing the temperature provides more thermal energy for hopping transport[10,28,95,96]. In this study, we measured the Drude mobilities at different temperatures ranging from 120 to 300 K. Samples were prepared in the *van der Pauw* configuration, with gold contacts evaporated on the four corners.

Fig. 3b shows the measured temperature-dependent mobility of $CuSbSe_2$ films. A clear decrease in mobility with an increase in temperature was observed, which is consistent with the behaviour of large polarons. At room temperature (300 K), Hall effect measurements gave a mobility of $0.93\pm0.05$ cm$^2\cdot$V$^{-1}\cdot$s$^{-1}$, while the mobility value extracted from the initial photoconductivity of OPTP measurements was $4.7\pm0.2$ cm$^2\cdot$V$^{-1}\cdot$s$^{-1}$. This difference in mobility values can be attributed to the different length scales of Hall effect and OPTP measurements. Hall effect measurements investigates charge carrier transport throughout the whole sample, while the mobility extracted from the initial OPTP signal represents the transport within a shorter range, usually well within one grain.

### *Theoretical insights into charge-carrier-phonon coupling in CuSbSe$_2$*

Having experimentally demonstrated an absence of charge-carrier localization in CuSbSe$_2$, which is unusual compared to most recently-investigated pnictogen-based perovskite-inspired materials[10,12,24,25], we aim now to establish the underlying factors enabling this behaviour. The strength of coupling between charge-carriers and acoustic phonons can be described by the acoustic coupling constant of Toyozawa *et al.*[98], $g_{ac}$, which is given by Eq. 1:

$$g_{ac} = \frac{E_d^2}{Ca_0} \cdot \frac{m}{3\pi\hbar^2} \qquad (1)$$

where $E_d$ is the acoustic deformation potential, $C$ the elastic constant, $m$ the mass of the charge-carrier considered and $\hbar$ the reduced Planck's constant. For values much less than one, we do not expect localization due to acoustic coupling. Charge-carrier localization can be expected even in stiff materials if they have large deformation potentials, due to the square proportionality seen in Eq. 1. The acoustic deformation potential describes the change in band edge position as we apply strain to a structure, and is described by Eq. 2:

$$E_d^{nk} = \frac{\delta \mathcal{E}_{nk}}{\delta S_{\alpha\beta}} \qquad (2)$$

where $\mathcal{E}_{nk}$ is the energy of band $n$ at wavevector $\mathbf{k}$, and $\boldsymbol{S}_{\alpha\beta}$ is the uniform stress tensor[97]. Table 1 lists the calculated acoustic deformation potentials along the different directions of the CuSbSe$_2$ crystal. The average values for CuSbSe$_2$ (1.69 eV for VBM; 6.51 eV for CBM) are much lower than those of Cs$_2$AgBiBr$_6$ (13.7 eV for VBM; 14.7 eV for CBM), which undergoes charge-carrier localization, and comparable to the values of CsPbBr$_3$ (2.2 eV for VBM; 6.3 eV for CBM), which has delocalized charge-carriers[10]. The low acoustic deformation potential is consistent with polarons being large in CuSbSe$_2$.

Deformation potentials are typically calculated by comparing a series of statically strained unit cells to the ground state. This is because in traditional zincblende tetrahedral semiconductors, there are no internal structural degrees of freedom. However, we also need to investigate how allowing relaxation of internal coordinates would affect these computations. In the case of CuSbSe$_2$, allowing atomic relaxation results in a reduction in the predicted deformation potentials by as much as 59% ($E_{d,cc}^{CBM}$) as compared to the static case (Supplementary Fig. 6a). To probe the cause of this reduction, we compared cation-anion bond lengths before and after atomic relaxation. We found that when the unit cell was strained along the *c*-axis and allowed to relax, there were comparatively little changes in intralayer bond lengths because these distortions were mostly taken up by an increase in interlayer spacing (Fig. 4b). Referring to Figs. 2b and 4c, we see that the band edges are dominated by intralayer bonding, and so relaxation of these bonds towards their ground state configuration should minimize changes in the electronic structure, thus minimising the deformation potentials.

**Table 1 | Calculated properties related to carrier-phonon coupling in CuSbSe$_2$ along different principal axes.** $a_o$: lattice parameter; $E_d^{VBM}$: acoustic deformation potential of the valence band maximum; $E_d^{CBM}$: acoustic deformation potential of the conduction band minimum; $g_{ac}$: acoustic coupling constant; $C_{iii}$: Diagonal component of the elastic tensor $\epsilon_\infty$: dielectric constant at high frequency; $\epsilon_{stat}$: static dielectric constant; $m_h^*$: effective mass of holes (related to electronic conductivity); $m_h^*$: effective mass of electrons (related to electronic conductivity); $\alpha_h$: Fröhlich coupling constant of holes; $\alpha_e$: Fröhlich coupling constant of electrons. $E_b$: Wannier-Mott binding energies

|  | *a* | *b* | *c* | Average |
|---|---|---|---|---|
| $a_o$ (Å) | 6.457 | 4.034 | 14.929 | |
| $E_d^{VBM}$(eV) | 1.16 | 1.79 | 2.11 | 1.69 |
| $E_d^{CBM}$ (eV) | 6.60 | 6.32 | 6.62 | 6.51 |
| $C_{iii}$ $(GPa)$ | 75.5 | 81.7 | 60.4 | 72.6 |
| $g_{ac}^{VBM}$ | 0.001 | 0.003 | 0.003 | 0.002 |
| $g_{ac}^{CBM}$ | 0.007 | 0.010 | 0.010 | 0.090 |
| $\epsilon_\infty$ | 10.1 | 12.5 | 11.4 | 11.3 |
| $\epsilon_{stat}$ | 12.0 | 40.4 | 16.5 | 23.0 |
| $m_h^*$ | 1.44 | 1.30 | 2.38 | 1.60 |
| $m_e^*$ | 0.29 | 0.41 | 0.94 | 0.43 |

| $\alpha_h$ | 0.55 | 1.77 | 1.17 | 1.59 |
| $\alpha_e$ | 0.25 | 0.99 | 0.73 | 0.82 |
| $E_b$ (meV) | 22.3 | 2.6 | 33.6 | 8.7 |

The strength of coupling between charge-carriers and longitudinal optical (LO) phonons is described by the Fröhlich coupling constant, $\alpha$, given by Eq. 3.

$$\alpha = \frac{e^2}{4\pi\epsilon_0}\Big(\frac{1}{\epsilon_\infty} - \frac{1}{\epsilon_{stat}}\Big)\sqrt{\frac{m^*}{2\omega_{LO}\hbar^3}} \qquad (3)$$

In Eq. 3, $\epsilon_0$ is the vacuum permittivity while $\epsilon_\infty$ and $\epsilon_{stat}$ are the calculated optical and static dielectric constants, respectively. $m^*$ is the (conductivity) effective mass of the charge-carrier considered, while $\omega_{LO}$ is the effective longitudinal optical (LO) phonon frequency, and $\hbar$ is the reduced Planck's constant. The values of these properties are shown in Table 1. $\omega_{LO}$ is 138 cm$^{-1}$, and was calculated as an average over all Γ-point modes weighted by the dipole moment they produce (since Fröhlich coupling arises due to interactions between charge-carriers and optical phonon modes producing local dipoles)[97]. The average Fröhlich coupling constants of holes and electrons ($\alpha_h$ = 1.59, $\alpha_e$ = 0.82) are both in the weak regime, lower than those found in ABZ$_2$ materials like NaBiS$_2$ ($\alpha_h$ = 2.92, $\alpha_e$ = 1.40)[12], AgBiS$_2$ ($\alpha_h$ = 1.63, $\alpha_e$ = 1.09)[14], as well as methylammonium lead iodide perovskites (2-3)[24]. The low Fröhlich coupling constants are well below the range typically considered to be strong[99], showing that carrier localization due to coupling with LO phonons should not occur in CuSbSe$_2$.

**Discussion**

***Mechanisms for weak charge-carrier-phonon coupling in CuSbSe$_2$***

In light of the results presented, we can begin to understand why strong carrier localization is not seen in CuSbSe$_2$ in terms of the structure and bonding present in the material. As mentioned previously, the low acoustic deformation potential $E_d$ of CuSbSe$_2$ is the prime factor in causing weak acoustic coupling. Since $E_d$ describes the change in band edge positions when the lattice is distorted, which is in turn influenced by the nature of bonding between atoms, analysis of the bonding environment can provide insights into the magnitude of the acoustic deformation potential.

The electronic structure of CuSbSe$_2$ is shown in the middle panel of Fig. 2b, with the orbital-projected density of states (pDOS) and band structure diagrams shown. On the far right of the same panel are crystal orbital Hamilton population (COHP) diagrams for Sb-Se and Cu-Se bonds, which allow us to determine whether these are bonding, anti-bonding or non-bonding interactions. With –COHP set as the horizontal axis, positive values represent bonding interactions, negative values indicate anti-bonding interactions between atoms, while values near the centre line indicate non-bonding interactions. The more positive –COHP value of the Sb-Se bond reveals stronger covalent bonding between Sb and Se atoms compared to Cu and Se atoms. We see the VBM of CuSbSe$_2$ mainly consists of Cu(d)-Se(p) anti-bonding states, with much weaker contributions from Sb(s,p)-Se(p), which are approximately non-bonding . These non-bonding states are the result of Sb s-p orbital mixing, which occurs through a second order Jahn-Teller distortion, resulting in a familiar 'non-bonding' lone-pair projected into the interlayer space.

From Table 1, it can be seen that there is more anisotropy in the deformation potentials at the VBM than CBM. This anisotropy in the VBM deformation potentials can be understood by considering the inequivalent distortions of the Cu-Se tetrahedra caused by strains along the principal crystallographic axes. Whilst all Cu-Se bonds are equivalent in both bond length and strength, these tetrahedra are arbitrarily rotated with respect to the principal axes. Thus, straining along these principal axes will cause inequivalent changes in the electronic structure at the VBM, leading to differences in the deformation potentials. For example, strain along the *a*-axis causes scissoring of pairs of Cu-Se bonds rather than significant changes in bond length, whereas strain along the *b* axis distorts all four bonds.

The magnitude of the deformation potential is substantially reduced by structural relaxation in this flexible crystal structure. We posit that quasi-2D structures with interlayer void space exhibits reduced deformation potentials compared to 3D structures, as strain-induced changes in bond length can be compensated by modulating the interlayer-spacing. To demonstrate this, we analyzed the change in cation-anion bond lengths (after relaxation) as the $CuSbSe_2$ lattice was strained along the *c*-axis (*i.e.*, perpendicular to $CuSbSe_2$ layers). The bonds considered are highlighted in Fig. 4a, and listed with the corresponding colours in Fig. 4b. When the strain reached $\pm 5\%$, the changes in most bond lengths were below 1%, and the maximum change (Cu1-Se3 bond) was only around 2%. This is not explained by misalignment of the strain to bonding vectors, as we see differences of more than 4% for the same Cu1-Se3 bond in the unrelaxed case (i.e., for a uniform distribution of strain along the inter-atomic distances), and is also in contrast to the large change in interlayer distance of $\pm 20\%$ under $\pm 5\%$ *c*-axis strain. The fact that Cu-Se bonds exhibit more changes than Sb-Se bonds agrees with the COHP calculation results that indicate that the Cu-Se bonds are overall weaker due to the filled antibonding states in the VBM[100]. The strong relaxation in Cu-Se and Sb-Se bond lengths correlates well with the general reduction in acoustic deformation potentials presented in Supplementary Fig. 6a, and suggests deformation potentials in Table 1, while low in absolute terms, are themselves an overestimation. This phenomenon should be considered when calculating deformation potentials in complex materials with similar structures to BiOI[39] and $CuSbSe_2$.

We also performed COHP calculations for the interlayer Sb-Se bonding interaction and derived an integrated crystal orbital Hamilton population (ICOHP) value as a measure of the covalent bonding strength (Fig. 4c). The much higher values of intralayer Sb-Se bonds (labelled in-layer in Fig. 4c) than interlayer Sb-Se bonds indicate that the interlayer covalent interaction is significantly weaker than the intralayer case, which is consistent with considering CuSbSe$_2$ as a layered material.

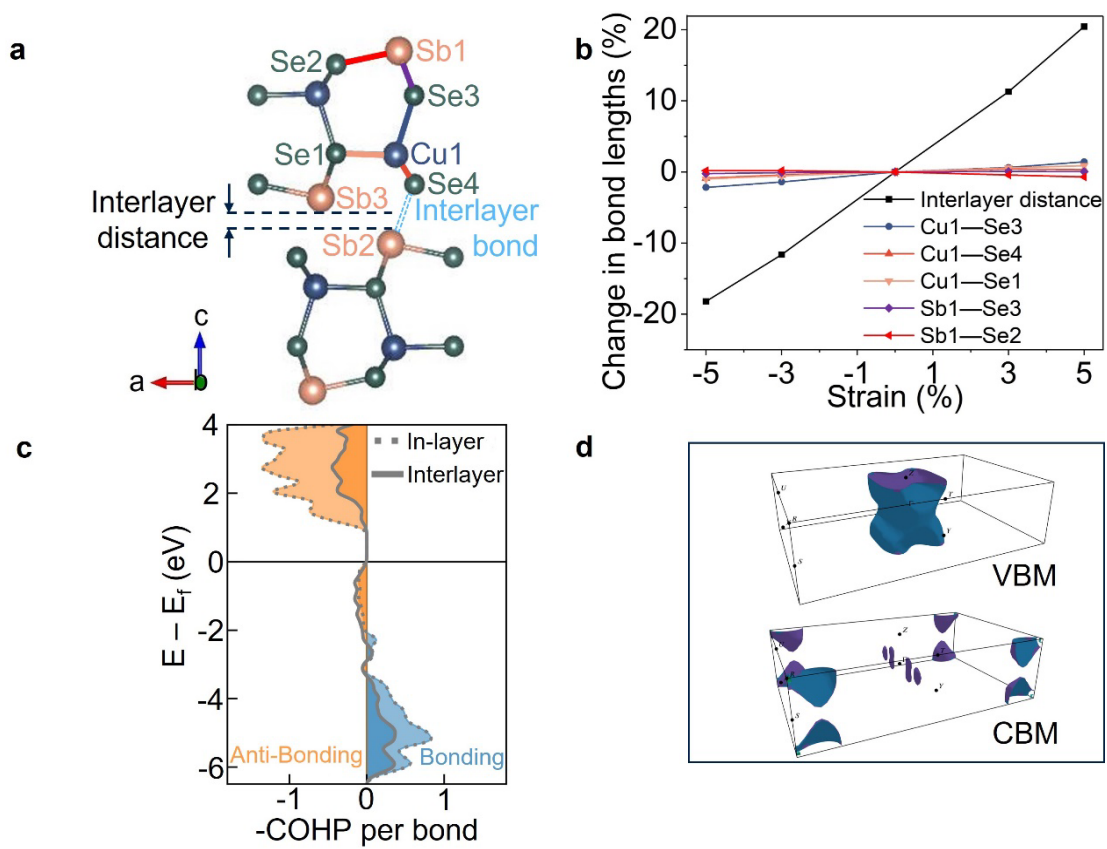

**Fig. 4 | Computational analysis of CuSbSe$_2$. a,** Structure of CuSbSe$_2$, with key atoms labelled, and the interlayer distance defined as the perpendicular distance between Sb2 and Sb3. **b,** Percentage changes in bond lengths and interlayer distance of CuSbSe$_2$ as a function of strain along the *c*-axis. All calculated bond lengths shown are after relaxation of the atoms in the structure after distortion. A disproportionally large change

in the interlayer distance is observed as compared to bond lengths for a given strain. **c,** Calculated crystal orbital Hamilton population (COHP) per bond of in-layer (dash line) and interlayer (solid line) Sb-Se bonds. The bonding and anti-bonding interactions are represented by blue and orange, respectively. **d,** Fermi iso-surface 0.1 eV below the VBM (top figure) and above the CBM (bottom figures).

Electronic dimensionality also has an important effect on charge-carrier-phonon coupling, which can be described to a first approximation by the continuum model of Toyozawa, which considers both acoustic and optical phonon fields[98,101]. A 3D electronic structure can be advantageous by having an energy barrier against charge-carrier localization, but the energy barrier height should also be accounted for[28,102]. As for electronically 2D materials, the tendency to undergo strong coupling to acoustic phonons depends on the acoustic coupling factor $g_{ac}$. When $g_{ac} > 1$, barrierless charge-carrier localization is energetically favorable. On the contrary, for $g_{ac} < 1$, charge-carrier localization should not occur because the lattice energy increases as charge-carriers become more localized[102]. For 1D materials, this model predicts spontaneous localization in all cases. The electronic dimensionality of a semiconductor may be probed by analyzing the Fermi surfaces slightly above and below the CBM and VBM, which are shown for $CuSbSe_2$ in Fig. 4d. These surfaces are representative of the states occupied by free charge-carriers in the material as a result of thermal or optical excitation. Planar or columnar motifs are indicative of 1D and 2D structures respectively, indicating weak dispersion along the flat planar/axial direction(s).

Meanwhile, ellipsoidal (closed-surface) motifs show dispersion in all directions and so are hallmarks of 3D electronic structures[66]. The VBM of CuSbSe$_2$ is unambiguously 2D in this transport regime due to the presence of a single columnar surface showing weak dispersion along the $c$-axis, while the CBM shows a number of ellipsoidal and closed rod-like structures, suggestive of an electronic structure that is 3D or close to 3D (Fig. 4d). The near-3D nature of the CBM is consistent with the lower CB being dominated by Sb-Se antibonding states, and there being weak interactions between the Sb and Se species across the interlayer gaps. By contrast, the 2D nature of the VBM is consistent with Cu-Se interactions, which dominate the upper VB, mostly occurring within each layer. The combination of the relatively high electronic dimensionality (especially in the CBM), and low $g_{ac}$ values overall are consistent with the band-like transport in CuSbSe$_2$.

This deviation in the electronic dimensionality from the structural dimensionality in CuSbSe$_2$ is consistent with what has been found in other pnictogen-based semiconductors. For example, although Cs$_2$AgBiBr$_6$ has a 3D crystal structure, its electronic dimensionality is significantly lower[103], which is one of the factors contributing to carrier localization in this material. As another example, although Sb$_2$Se$_3$ and Sb$_2$S$_3$ both have the same quasi-1D crystal structure, we see a 2D VBM in the former and a 3D VBM in the latter[66]. In the case of CuSbSe$_2$, although the electronic dimensionality of the VBM matches its quasi-2D structural dimensionality, the CBM has 3D-like character. This emphasizes the importance of evaluating the electronic dimensionality and considering other important properties, such as the acoustic and Fröhlich coupling constants, in order to rationalize the nature of carrier-phonon interactions.

Another feature of CuSbSe$_2$ is its weak Fröhlich interaction, which primarily arises due to the small difference between the electronic and static dielectric constants, $\epsilon_\infty$ and $\epsilon_{\text{stat}}$ (refer to Table 1 and Eq. 3). This occurs when the ionic dielectric contribution is low relative to electronic contributions, and occurs due to both i) a high electronic dielectric contribution along all principal axes, and ii) a low ionic dielectric contribution, especially along the $a$- and $c$-axes (Table 1). The high electronic contribution is due to the small bandgap (since $\epsilon_\infty \propto E_{\text{g}}^{-0.5}$). In addition, the high density of states near the band-edges will lead to a stronger interaction between electrons and light, giving a higher refractive index and higher $\epsilon_\infty$ (since $\epsilon_\infty \propto n^{0.5}$, where $n$ is the refractive index[104]). To understand the cause of the low ionic dielectric contribution, we calculated the Born effective charge (BEC) tensors for the different sublattices in CuSbSe$_2$ (Table 2). The Born effective charges ($Z_{\alpha,ij}^*$), also known as dynamical charges, describe the change in polarization in direction $i$ when the sublattice of atoms ($\alpha$) is displaced along direction $j$[105,106],

$$Z^*_{\alpha,ij} = \frac{\partial P_i}{\partial u_{\alpha,j}}.$$ (4)

For materials with strong ionic-covalent bonding, the Born effective charges can be significantly larger than the formal oxidation states. The BECs for Cu1 are close to the oxidation state of the species, however, when considering the whole BEC tensor for Sb1 and Se1 atoms, we observe BEC values higher than the formal oxidation states with displacements along the $b$ direction, while net BECs (summing over columns) for displacements in the $a$ and $c$ directions are lower than the oxidation states. The low net dynamical charges of Sb1 and Se1 for $a$ and $c$ displacements are reflected by their close $\epsilon_\infty$ and $\epsilon_{\text{stat}}$ values, and contribute to the low Fröhlich coupling constants in these directions (Table 1). The anomalously large contributions of the Sb1 and Se1 atoms along the $b$-axis are of interest. These can be explained by either a change in polarization of Sb1 and Se1 atoms upon displacement, or a direct transfer of charge between the two species, however it is difficult to say for sure without further investigation[106]. The lone-pair on the Sb1 atom, and its origins in a symmetry breaking interaction between hybridized Sb-s, p orbitals and Se p orbitals[67] can explain this. Changes in the symmetry (*e.g.* via sub-lattice displacement) of the Sb-Se coordination sphere will change how the lone pair is expressed, leading to strong deviations in BEC.

Nevertheless, despite the larger ionic contributions to the dielectric constant along the

*b*-axis, Fröhlich coupling constants remain <2 (Table 1).

**Scheme 1 | Calculated Born effective charge and oxidation state of atoms in CuSbSe₂,** with the net BEC values for displacements along principal crystallographic axes in square brackets.

$$Z^*_{Cu1(I)} = \begin{pmatrix} 0.87 & 0 & -0.07 \\ 0 & 1.23 & 0 \\ 0.28 & 0 & 1.18 \end{pmatrix} Z^*_{Sb1(III)} = \begin{pmatrix} 2.53 & 0 & -0.62 \\ 0 & 5.65 & 0 \\ -1.50 & 0 & 2.06 \end{pmatrix}$$

$$[1.14 \quad 1.23 \quad 1.12] \qquad\qquad [1.03 \quad 5.65 \quad 1.44]$$

$$Z^*_{Se1(ii)} = \begin{pmatrix} -1.06 & 0 & -0.21 \\ 0 & -4.18 & 0 \\ -0.45 & 0 & -1.08 \end{pmatrix} Z^*_{Se2(ii)} = \begin{pmatrix} -2.34 & 0 & 0.04 \\ 0 & -2.71 & 0 \\ 0.67 & 0 & -2.16 \end{pmatrix}$$

$$[-1.51 \quad -4.18 \quad -1.29] \qquad\qquad [-1.67 \quad -2.71 \quad -2.12]$$

**Conclusion**

In conclusion, we have found $CuSbSe_2$ to be a heavy pnictogen-based chalcogenide that can avoid charge-carrier localization, which we determined through a combination of experiment and computations. A novel thiol-amine solution processing method was employed to achieve phase-pure $CuSbSe_2$ thin films. OPTP measurements on $CuSbSe_2$ revealed a photoconductivity decay timescale of approximately 50 ps, substantially slower than if carrier localization were present. Temperature-dependent Hall effect measurements confirmed the presence of large polarons based on the decrease in mobility with increases in temperature. Through DFT calculations, we found that both the acoustic and Fröhlich coupling constants are lower than those of many other heavy pnictogen-based materials, which support the finding that $CuSbSe_2$ has weaker charge-carrier-phonon coupling. To reveal the underlying mechanisms enabling $CuSbSe_2$ to avoid charge-carrier localization, key parameters, including the bonding/anti-bonding nature of the crystal orbitals at the band extrema, and changes in bond lengths and interlayer spacing as a function of distortions, as well as the Born effective charges of ions, were investigated via computational tools. It was shown that the low deformation potentials and relatively high electronic dimensionality contribute to $g_{ac} < 1$, hence an absence of strong coupling to acoustic phonons. Meanwhile, the weak Fröhlich coupling is due to the high electronic contribution (mostly due to the small bandgap) and low ionic contribution to the dielectric constants. The latter arises from the Born effective charges of Sb, Cu and Se not substantially deviating from their formal oxidation states (in contrast to lead-halide perovskites)[107]. Based on our investigations, we propose that the free volumes (*e.g.* interlayer gaps) in the lattice can help to minimize the effect of lattice distortions on the bonding environment and lower the deformation potential. At the same time, electronic coupling across the interlayer gap between species contributing to the band-edge density of states can increase the electronic dimensionality, which reduces the likelihood of self-trapping. Finally, materials with low ionic contributions to the dielectric constant are desired to minimize Fröhlich coupling, but this needs to be balanced with the effect on the capture cross-section of charged defects. These insights are valuable for the future design of solar absorbers that have band-like transport.

**Data Availability**

The raw experimental data generated in this paper and the Supplementary Information can be found from the Oxford University Research Archive (ORA) Data Repository, under the accession number [to be completed later].

**Code Availability**

The computational files associate with the calculations made in this paper can be found from [to be completed later].

**Acknowledgements**

Y.F. and H.L. thank Xinwei Wang and Alex Ganose for valuable discussions. Y.F., H.L. and R.L.Z.H. thank the UK Research and Innovation for funding through a Frontier Grant (no. EP/X022900/1), awarded via the European Research Council 2021 Starting Grant scheme. Y.-T.H. and R.L.Z.H thank the Engineering and Physical Sciences Research Council (EPSRC) for funding (no. EP/V014498/2). Y.F., H.L., Y.-T.H. and R.L.Z.H. thank the Henry Royce Institute for support through the Industrial Collaboration Programme, funded through EPSRC (no. EP/X527257/1). L.M.H and M.R thank the Engineering and Physical Sciences Research Council (EPSRC) for funding. R.L.Z.H. also thanks the Royal Academy of Engineering for funding via the Research Fellowships scheme (no. RF\201718\17101).

**Author Contributions**

R.L.Z.H. conceived of and supervised this project as a whole. Y.F. developed the method for synthesizing $CuSbSe_2$ by solution processing and performed X-ray diffraction, Raman spectroscopy, Fourier-transform infrared spectroscopy, UV-visible spectroscopy and room-temperature Hall effect measurements, supervised by R.L.Z.H.

H.L. performed the computations, with support from S.R.K. and Y.W.W, and supervised by A.W. M.R. performed the optical pump terahertz probe spectroscopy measurements and Elliot model fit, supervised by L.M.H. Y.-T.H. performed transient absorption spectroscopy, supervised by R.L.Z.H and A.R. C.-W.C performed low-temperature Hall effect measurements, supervised by B.A.P. S.J.Z performed photothermal deflection spectroscopy. H.D helped optimize the annealing of $CuSbSe_2$ thin films, supervised by S.H. M.A.M provided supports on thin film deposition equipment. All authors contributed to writing the manuscript.

## Competing Interests

The authors declare no competing financial or non-financial interests.

## Additional information

**Supplementary information** is available for this paper at XXX

Correspondence and requests for materials should be addressed to R.L.Z.H.